\documentclass[english,10pt,conference]{IEEEtran}
\usepackage[T1]{fontenc}
\usepackage[latin9]{inputenc}
\usepackage[letterpaper]{geometry}
\usepackage{color}
\usepackage{float}
\usepackage{amsmath}
\usepackage{graphicx}
\usepackage{amssymb}

\makeatletter

\floatstyle{ruled}
\newfloat{algorithm}{tbp}{loa}
\floatname{algorithm}{Algorithm}

\usepackage{tikz}

\IEEEoverridecommandlockouts

\makeatother

\usepackage{babel}

\begin{document}

\title{IMP: A Message-Passing Algorithm\\
for Matrix Completion }

\author{\authorblockN{Byung-Hak Kim, Arvind Yedla, and Henry D. Pfister}\thanks{This material is based upon work supported by the National Science Foundation under Grant No. 0747470. Any opinions, findings, conclusions, or recommendations expressed in this material are those of the authors and do not necessarily reflect the views of the National Science Foundation.}
\authorblockA{Department of Electrical and Computer Engineering,
Texas A\&M University\\
 Email: \{bhkim,yarvind,hpfister\}@tamu.edu}}
\maketitle
\begin{abstract}
A new message-passing (MP) method is considered for the matrix completion
problem associated with recommender systems. We attack the problem
using a (generative) factor graph model that is related to a probabilistic
low-rank matrix factorization. Based on the model, we propose a new
algorithm, termed IMP, for the recovery of a data matrix from incomplete
observations. The algorithm is based on a clustering followed by inference
via MP (IMP). The algorithm is compared with a number of other matrix
completion algorithms on real collaborative filtering (e.g., Netflix)
data matrices. Our results show that, while many methods perform similarly
with a large number of revealed entries, the IMP algorithm outperforms
all others when the fraction of observed entries is small. This is
helpful because it reduces the well-known cold-start problem associated
with collaborative filtering (CF) systems in practice. \vspace{-2mm}
\end{abstract}

\section{Introduction \label{sec:Intro}\vspace{0.5mm}}

An important new inference problem, called the \emph{matrix completion}
problem, has recently come to light; it combines many elements of
compressed sensing and collaborative filtering. This problem involves
the recovery of a data matrix from incomplete (or corrupted) information
and is of great practical interest over a wide range of fields\emph{
}\cite{Candes-08}. The basic idea is summarized well in the following
quote:
\begin{quote}
\emph{{}``In its simplest form, the problem is to recover a matrix
from a small sample of its entries, and comes up in many areas of
science and engineering including collaborative filtering, machine
learning, control, remote sensing, and computer vision... Imagine
now that we only observe a few entries of a data matrix. Then is it
possible to accurately\textemdash{}or even exactly\textemdash{}guess
the entries that we have not seen?'' -} Candes and Plan \cite{Candes-09}
\end{quote}
In the Netflix challenge, for example, one is given a subset of large
data matrix in which rows are users and columns are movies (e.g.,
see the Netflix Prize \cite{Netflix-web}). An overwhelming portion
of the user-movie matrix (e.g., 99\%) is unknown and the observation
matrix is very sparse because most users rate only a few movies. Randomness
in the ratings process implies that one can also interpret the ratings
as noisy observations of some true matrix. 

The goal is to predict the rating that a user would give, to a movie
he/she has not watched, based on the observed ratings. In other words,
the problem is to recover missing ratings of a data matrix using the
subset of observed movie ratings. In general, it would seem that this
problem is difficult, if not impossible. However, if the unknown matrix
has some structure, then approximate recovery is possible. Recent
progress on the matrix completion problem can be largely divided into
two areas: 
\begin{enumerate}
\item The first area considers efficient models and practical algorithms.
For the matrix completion problem, many researchers use models based
on the assumption that the matrix has low rank. This assumption allows
one to reformulate the problem into rank (or nuclear norm) minimization
problem under certain incoherence assumptions \cite{Candes-08}. For
exact and approximate matrix completion, these models lead to convex
relaxations, and semi-definite programming (SDP) \cite{Cai-arxiv08}\cite{Lee-arxiv09}\cite{Keshavan-arxiv09b}\cite{Dai-arxiv09},
and Bayesian-based approaches \cite{Salakhutdinov-icml08}. \vspace{0.25mm}
\item The second area involves exploration of the fundamental limits of
these methods. Prior work has developed some precise relationships
between sparse observation models and the recovery of missing entries
under the restriction of low-rank matrices or clustering models \cite{Candes-09}\cite{Candes-arxiv09}\cite{Keshavan-arxiv09a}\cite{Aditya-itsub09}\cite{Vishwanath-arxiv10}.
This area is closely related with the practical issues known as the
cold-start problem of the recommender system \cite{Schein-02}. That
is, giving recommendations to new users who have submitted only a
few ratings, or recommending new items that received only a few ratings
from users. In other words, how many ratings are needed to generate
good recommendations? 
\end{enumerate}
Unlike this prior work, this paper considers an important subclass
of the matrix completion problem where the entries (drawn from a finite
alphabet) are modeled by a (generative) factor graph. Based on this
factor graph model, we propose a MP based algorithm, termed IMP, to
estimate missing entries. This algorithm seems to share some of the
desirable properties demonstrated by MP in its successful application
to modern coding theory \cite{Gallager-1963}. The IMP algorithm tries
to combine the benefits of soft clustering of users/movies into groups
and message-passing based on the unknown groups to make predictions.
In addition, simulation results for cold-start settings (i.e., less
than 0.5\% randomly sampled entries) show that the cold start problem
is reduced greatly by IMP in comparison to other methods on real collaborative
filtering (or Netflix) data matrices. 

The paper is structured as follows. After defining the factor graph
model in Section \ref{sec:FGM}, we introduce the IMP algorithm in
Section \ref{sec:IMP}. In Section \ref{sec:Simulations}, we discuss
the algorithm performance via experimental results, and give conclusions
in Section \ref{sec:Concl}.\vspace{-2mm}

\section{Factor Graph Model \label{sec:FGM}\vspace{0mm}}

\begin{figure}[t]
\begin{minipage}[c][1\totalheight][t]{0.45\textwidth}%
{\small \centering 
\begin{tikzpicture}[>=stealth,scale=0.53]
\draw (-1,2) rectangle +(13,1);
\draw (5.5,2.5) node {permutation};
\draw (-1,5) rectangle +(13,1);
\draw (5.5,5.5) node {permutation};
\foreach \x/\userlabel/\movielabel in {0/0/0,1/1/1,2/2/2,3/3/3,4/4/4,5/5/5,6/6/6,11/N/M}
{
\filldraw[gray] (\x,1)+(2pt,2pt) circle (2pt) node[black,below=3pt] {$U_{\userlabel}$};
\draw (\x,1)+(2pt,4pt) -- (\x,2);
\draw (\x,1)+(2pt,4pt) -- ([xshift = 4pt]\x,2);
\draw (\x,1)+(2pt,4pt) -- ([xshift = 2pt]\x,2);
\draw (\x,1)+(2pt,4pt) -- ([xshift = -2pt]\x,2);
\draw (\x,1)+(2pt,4pt) -- ([xshift = 6pt]\x,2);

\draw[thick,gray] (\x,3)+(0.5cm,0pt) -- ([xshift=0.5cm]\x,5);

\filldraw (\x,7)+(2pt,2pt) circle (2pt) node[black,above=3pt] {$V_{\movielabel}$};
\draw (\x,7)+(2pt,0pt) -- (\x,6);
\draw (\x,7)+(2pt,0pt) -- ([xshift = 4pt]\x,6);
\draw (\x,7)+(2pt,0pt) -- ([xshift = 2pt]\x,6);
\draw (\x,7)+(2pt,0pt) -- ([xshift = -2pt]\x,6);
\draw (\x,7)+(2pt,0pt) -- ([xshift = 6pt]\x,6);
}

\draw[thick,gray] (-1,3)+(0.5cm,0pt) -- ([xshift=0.5cm]-1,5);

\foreach \x in {0.425,1.425,2.425,...,6.425,11.425} {
	\filldraw (\x,3.925) rectangle +(4pt,4pt);
}
\filldraw (-0.575,3.925) rectangle +(4pt,4pt);
\foreach \x in {7.5,8,8.5,...,10} {
\foreach \y in {1,4,7} {
\filldraw (\x,\y) circle (1pt);
}}

\draw (-2,1.5) node {$\scriptstyle\mathbf{U}$};
\draw (13,1.5) node {$\scriptstyle\mathbf{y}^{(i)}$};
\draw[->,thick] (13,1.8) -- (13,2.4);
\draw (-2,1) node {\scriptsize{Users}};
\draw (-2,4.25) node {\scriptsize{Ratings}};
\draw (-2,3.75) node {$\scriptstyle\mathbf{R_{O}}$};
\draw (-2,6.5) node {$\scriptstyle\mathbf{V}$};
\draw (13,6.5) node {$\scriptstyle\mathbf{x}^{(i)}$};
\draw[->,thick] (13,6.2) -- (13,5.6);
\draw (-2,7) node {\scriptsize{Movies}};
\end{tikzpicture}

}%
\end{minipage}\vspace{-2mm}

\caption{\label{fig:tanner}The factor graph model for the matrix completion
problem. The graph is sparse when there are few ratings. Edges represent
random variables and nodes represent local probabilities. The node
probability associated with the ratings implies that each rating depends
only on the movie group (top edge) and the user group (bottom edge).
Synthetic data can be generated by picking i.i.d. random user/movie
groups and then using random permutations to associate groups with
ratings. Note $\mathbf{x}^{(i)}$ and $\mathbf{y}^{(i)}$ are the
messages from movie to user and user to movie during iteration $i$
for the Algorithm \ref{alg:IMP}.\vspace{0mm}}

\end{figure}
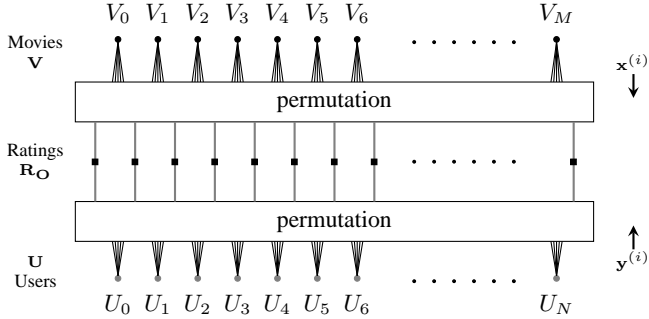
Consider a collection of $N$ users and $M$ movies when the set $O$
of user-movie pairs have been observed. The main theoretical question
is, {}``How large should the size of $O$ be to estimate the unknown
ratings within some distortion $\delta$?''. Answers to this question
certainly require some assumptions about the movie rating process.
So we begin differently by introducing a probabilistic model for the
movie ratings. The basic idea is that \emph{hidden} variables are
introduced for users and movies, and that the movie ratings are conditionally
independent given these hidden variables. It is convenient to think
of the hidden variable for any user (or movie) as the \emph{user group}
(or \emph{movie group}) of that user (or movie) and this can be viewed
as a simplistic assumption about the psychological nature of movie
preferences \cite{Backstrom-kdd06}\cite{crandall-kdd08}. In this
context, the rating associated with a user-movie pair depends only
on the user group and the movie group. 

Since the number of movie groups are very small compared to the number
of movies, this idea is similar to mapping movies to a low-dimensional
movie group. Each movie group may correspond to a genre (e.g., comedy,
drama, action, ...). Each user group tries to capture sets of users
that have similar taste in movies. For example, a movie may be classified
as a comedy, and a user may be classified as a comedy lover. The model
may use 20 to 40 such groups to locate each movie and user in a multidimensional
space. It then predicts a user\textquoteright{}s rating of a movie
according to the movie\textquoteright{}s rating on the dimensions
that person cares about most since similar user/movie map to similar
groups in the low-dimensional (group) space.

The goal is to design a probabilistic mapping such that reflects group
associations in the low-dimensional (group) space. Let there be $g_{u}$
user groups, $g_{v}$ movie groups, and define $[k]\triangleq\left\{ 1,2,\ldots,k\right\} $.
The user group of the $n$-th user, $U_{n}\in[g_{u}]$, is a discrete
random variable drawn from $\Pr(U_{n}=u)\triangleq p_{U}(u)$ and
$\mathbf{U}=U_{1},U_{2},\ldots,U_{N}$ is the user group vector. Likewise,
the movie group of the $m$-th movie, $V_{m}\in[g_{v}]$, is a discrete
random variable drawn from $\Pr(V_{m}=v)\triangleq p_{V}(v)$ and
$\mathbf{V}=V_{1},V_{2},\ldots,V_{M}$ is the movie group vector.
Then, the rating of the $m$-th movie by the $n$-th user is a discrete
random variable $R_{nm}\in\mathcal{R}$ (e.g., Netflix uses $\mathcal{R}=[5]$)
drawn from $\Pr(R_{nm}=r|U_{n}=u,V_{m}=v)\triangleq w(r|u,v)$ and
the rating $R_{nm}$ is \emph{conditionally independent} given the
user group $U_{n}$ and the movie group $V_{m}$. Let $\mathbf{R}$
denote the rating matrix and the observed submatrix be $\mathbf{R}_{O}$
with $O\subseteq[N]\times[M]$. In this setup, some of the entries
in the rating matrix are observed while others must be predicted.
The conditional independence assumption in the model implies that
\vspace{-.5mm} \[
\Pr\left(\mathbf{R}_{O}|\mathbf{U},\mathbf{V}\right)\triangleq\prod_{(n,m)\in O}w\left(R_{nm}|U_{n},V_{m}\right).\]

Specifically, we consider the factor graph (composed of 3 layers,
see Figure \ref{fig:tanner}) as a randomly chosen instance of this
problem based on this probabilistic model. The key assumptions are
that these layers separate the influence of user groups, movie groups,
and observed ratings. A random permutation is used to map the edges
attached to user nodes to the edges attached to movie nodes.

This model attempts to exploit correlation in the ratings based on
similarity between users (and movies). It also tries to include the
noisy rating process in the model and reduce the impact of corrupted
ratings on prediction by dimension reduction. These advantages allows
one to approximates real Netflix data generation process more closely
than other simpler factor models. In fact this model can be seen as
a generalization of \cite{Salakhutdinov-icml08} and \cite{Aditya-itsub09}.
It is also important to note that this is a \emph{probabilistic generative
model} which generalizes the clustering model in and also allows one
to evaluate different learning algorithms on synthetic data and compare
the results with theoretical bounds (see \cite{Kim-arxiv10} for details).\vspace{-1mm}

\section{The IMP Algorithm\label{sec:IMP}\vspace{0mm}}

\subsection{Initializing $w(r|u,v)$ for Group Ratings}

The IMP algorithm requires reasonable initial estimates, of the observation
model $w(r|u,v)$, to get started. To get these estimates, we cluster
users (and movies) first. The basic method uses a variable-dimension
vector quantization (VDVQ) clustering algorithm and the standard codebook
splitting approach known as the generalized Lloyd algorithm (GLA)
to generate codebooks whose size is any power of 2 \cite{Gersho-1992}.
Though our approach was motivated by the VDVQ clustering algorithm,
it turns out to be equivalent to soft $K$-means clustering with an
appropriate distance measure. So we will refer VDVQ clustering as
soft $K$-means clustering. 

The soft $K$-means clustering algorithm is based on the alternating
minimization of \textcolor{black}{the average distance between users
(or movies) and codebooks} (that contain no missing data). This leads
to alternating application of \emph{nearest neighbor} and \emph{centroid}
rules. The distance is computed only on the elements both vectors
share. In the case of users, one can think of this Algorithm \ref{alg:CR_alg}
as a {}``$K$-critics'' algorithms which tries to design $K$ critics
(i.e., people who have seen every movie) that cover the space of all
user tastes and each user is given a soft {}``degree of assignment
(or soft group membership)'' to each of the critics which can take
on values between 0 and 1. After soft-clustering users/movies each
into user/movie groups, we estimate $w(r|u,v)$ by computing the soft
frequency of each rating for each user-movie group pair. \vspace{-1mm}

\subsection{Message-Passing Updates of Group Vectors}

{\small }%
\begin{algorithm}[t]
{\small \caption{\label{alg:IMP}{\small IMP Algorithm}}
}{\small \par}

\textbf{\small Step I:}{\small{} Initialization of $w(r|u,v$) via
Algorithm \ref{alg:CR_alg} and randomized initialization of user/movie
group probabilities $\mathbf{x}_{m\to n}^{(0)}(v)$ and $\mathbf{y}_{n\to m}^{(0)}(u)$. }{\small \par}

\textbf{\small Step II:}{\small{} Recursive update for user/movie group
probabilities }\vspace{-2mm}

{\small \[
\mathbf{y}_{n\to m}^{(i+1)}(u)\!\propto\!\mathbf{y}_{n}^{(0)}(u)\!{\displaystyle \prod_{k\in\mathcal{V}_{n}\backslash m}}\!{\displaystyle \sum_{v}}w\left(r|u,v\right)\mathbf{x}_{k\to n}^{(i)}(v)\]
\[
\mathbf{x}_{m\to n}^{(i+1)}(v)\!\propto\!\mathbf{x}_{m}^{(0)}(v){\displaystyle \!\prod_{k\in\mathcal{U}_{m}\backslash n}}\!{\displaystyle \sum_{u}}w\left(r|u,v\right)\mathbf{y}_{k\to m}^{(i)}(u)\]
}{\small \par}

\textbf{\small Step III:}{\small{} Update $w(r|u,v)$ and output probability
of rating $R_{nm}$ given observed ratings }\vspace{-2mm}{\small{}
\[
{\textstyle \hat{p}_{R_{nm}|\mathbf{R}_{O}}^{(i+1)}(r)}\!\propto\!{\displaystyle \sum_{u,v}}\mathbf{y}_{n\to m}^{(i+1)}(u)w\left(r|u,v\right)\mathbf{x}_{m\to n}^{(i+1)}(v)\]
}\vspace{-2mm}
\end{algorithm}
\begin{figure*}[t]
\begin{centering}
\includegraphics[scale=0.31]{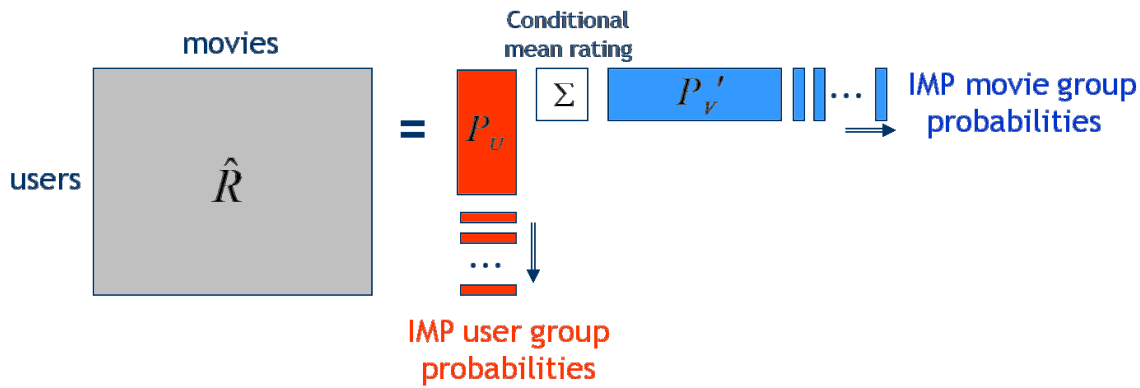}\vspace{-2mm}
\par\end{centering}

\caption{\label{fig:model}Minimum mean square estimator (MMSE) estimates $\hat{R}$
can be written as a matrix factorization. Each element of $\Sigma$
represents the conditional mean rating of $w\left(r|u,v\right)$ given
$u,\, v$ and each row of $P_{U}/P_{V}$ represents a user/movie group
probabilities. In contrast to the basic low-rank matrix model, we
add non-negativity (to $\Sigma,\, P_{U}$ and $P_{V}$) and normalization
constraints (to both $P_{U}$ and $P_{V}$).}
\vspace{-2mm}
\end{figure*}

\noindent Using the model from Section \ref{sec:FGM}, we describe
how message-passing can be used for the prediction of hidden variables
based on observed ratings. Ideally, we could perform exact inference
of our factor graph model. But exact learning and inference for this
model is intractable, so we turn to approximate message-passing algorithms
(e.g., the sum-product algorithm) \cite{Kschischang-it01}. The basic
idea is that the local neighborhood of any node in the factor graph
is tree-like (see \cite{Kim-arxiv10} for details). For iteration
$i$, we simplify notation by denoting the message from movie $m$
to user $n$ by $\mathbf{x}_{m\to n}^{(i)}$ and the message from
user $n$ to movie $m$ by $\mathbf{y}_{n\to m}^{(i)}$. The iteration
is initialized with \vspace{-2mm} \[
\mathbf{x}_{m\to n}(v)\!=\!\mathbf{x}_{m}(v)\!=\! p_{V}(v),\,\mathbf{y}_{n\to m}(u)\!=\!\mathbf{y}_{n}(u)\!=\! p_{U}(u).\]
The set of all users who rated movie $m$ is denoted $\mathcal{U}_{m}$
and the set of all movies whose rating by user $n$ was observed is
denoted $\mathcal{V}_{n}$. The exact update equations are given in
Algorithm \ref{alg:IMP}. The group probabilities are randomly initialized
by assuming that the initial group (of the user and movie) probabilities
are uniform across all groups.\vspace{-1mm}

\subsection{Approximate Matrix Completion}

\noindent {\small }%
\begin{algorithm}[t]
{\small \caption{\label{alg:CR_alg}{\small Initializing Group Ratings (shown only
for users)}}
}{\small \par}

\textbf{\small Step I:}{\small{} Initialization}\textcolor{black}{\small{} }{\small \par}

{\small ~~}\textcolor{black}{\small Let $i=j=0$ }{\small and $c_{m}^{(0,0)}(0)$}\textcolor{black}{\small ~be
the average rating vector of users for movie $m$.}{\small \par}

\textbf{\small Step II:}{\small{} Splitting of critics}{\small \par}

{\small ~~Set \[
c_{m}^{(i+1,j)}(u)\!=\!\begin{cases}
c_{m}^{(i,j)}(u) & u\!=\!0,\ldots,2^{i}\!-\!1\\
c_{m}^{(i,j)}(u\!-\!2^{i})\!+\! z_{m}^{(i+1,j)}(u) & u\!=\!2^{i},\ldots,2^{i+1}\!\!-\!1\end{cases}\]
where the $z_{m}^{(i+1,j)}(u)$ are i.i.d. random variables with small
variance.}{\small \par}

\textbf{\small Step III:}{\small{} Recursive soft $K$-means clustering
for $c_{m}^{(i,j)}(u)$ for $j=1,\,\ldots\,,\, J$.}{\small \par}

{\small ~~1. Each user is assigned a soft group membership }$\pi_{n}\left(u\right)${\small{}
to each of the critics using \[
\pi_{n}^{(i,j)}\left(u\right)\propto\mbox{exp}\left(-\beta\sqrt{\frac{1}{\left|\mathcal{V}_{n}\right|}\sum_{m\in\mathcal{V}_{n}}\left(c_{m,n}^{(i,j)}(u)-R_{nm}\right)^{2}}\right)\]
where $\mathcal{V}_{n}=\left\{ m\in\left[M\right]\,|\,(n,m)\in O\right\} $
and $g_{u}=2^{i+1}$. \vspace{1mm}}{\small \par}

{\small ~~2. Update all critics as}{\small \par}

{\small \[
c_{m}^{(i,j+1)}(u)\propto\sum_{n}\pi_{n}^{(i,j)}\left(u\right)c_{m}^{(i,j)}(u).\]
}{\small \par}

\textbf{\small Step IV:}{\small{} Repeat Steps II and III until the
desired number of critics }$g_{u}${\small{} is obtained. }{\small \par}

\textbf{\small Step V:}{\small{} Estimate of }$w(r|u,v)$

{\small ~~After clustering users/movies each into user/movie groups
with the soft group membership $\pi_{n}\left(u\right)$ and $\tilde{\pi}_{m}\left(v\right)$,
compute the soft frequencies of ratings for each user/movie group
pair as}{\small \par}

\[
w(r|u,v)\propto{\displaystyle \sum_{(n,m)\in O:R_{nm}=r}}\pi_{n}\left(u\right)\tilde{\pi}_{m}\left(v\right).\]
\vspace{-2mm}
\end{algorithm}
Since the primary goal is the prediction of hidden variables based
on observed ratings, the IMP algorithm focuses on estimating the distribution
of each hidden variable given the observed ratings. In particular,
the outputs of the algorithm (after $i$ iterations) are estimates
of the distributions for $R_{nm}$, $U_{n}$, and $V_{m}$. They are
denoted, respectively, as 

\noindent \begin{align*}
{\textstyle \hat{p}_{R_{nm}|\mathbf{R}_{O}}^{(i+1)}(r)}\! & \propto\!{\displaystyle \sum_{u,v}}\mathbf{y}_{n\to m}^{(i+1)}(u)w\left(r|u,v\right)\mathbf{x}_{m\to n}^{(i+1)}(v)\\
{\textstyle \hat{p}_{U_{n}|\mathbf{R}_{O}}^{(i+1)}(u)}\! & \propto\!\mathbf{y}_{n}^{(0)}(u)\!{\displaystyle \prod_{k\in\mathcal{V}_{n}}}\!{\displaystyle \sum_{v}}w\left(r|u,v\right)\mathbf{x}_{k\to n}^{(i)}(v)\\
\hat{p}_{V_{m}|\mathbf{R}_{O}}^{(i+1)}(v)\! & \propto\!\mathbf{x}_{m}^{(0)}(v){\displaystyle \!\prod_{k\in\mathcal{U}_{m}}}\!{\displaystyle \sum_{u}}w\left(r|u,v\right)\mathbf{y}_{k\to m}^{(i)}(u).\end{align*}
Using these, one can minimize various types of prediction error. For
example, minimizing the mean-squared prediction error results in the
conditional mean estimate (see Figure \ref{fig:model})\[
\hat{r}_{n,m}^{(i)}=\sum_{r\in\mathcal{R}}r\,{\textstyle \hat{p}_{R_{nm}|\mathbf{R}_{O}}^{(i)}(r)}.\]
\vspace{-2mm}

\subsection{Density Evolution (DE) Analysis \label{sec:DE_sec}}

\noindent DE is a well-known technique for analyzing probabilistic
message-passing inference algorithms that was originally developed
to analyze belief-propagation decoding of error-correcting codes and
was later extended to more general inference problems \cite{Montanari2007-arxiv07}.
It works by tracking the distribution of the messages passed on the
graph under the assumption that the local neighborhood of each node
is a tree. While this assumption is not rigorous, we consider that,
in Figure \ref{fig:tanner}, the outgoing edges from each user node
are attached to movie nodes via random permutations. This is identical
to the model used for irregular LDPC codes \cite{Richardson-it01}.
For this problem, the messages passed during inference consist of
belief functions for user groups (e.g., passed from movie nodes to
user nodes) and movie groups (e.g., passed form user nodes to movie
nodes). We have derived the DE equations for this problem and currently
in process of doing analysis based on them (see \cite{Kim-arxiv10}
for details). Like LDPC codes, we expect to see that the performance
of Algorithm \ref{alg:IMP} depends heavily on the degree structure
of the factor graph.\vspace{-1mm}

\section{Simulation Results with Real Data Matrices \label{sec:Simulations}\vspace{0mm}}

\subsection{Details of Training}

The key challenge of matrix completion problem is predicting the missing
ratings of a user for a given item based only on very few known ratings
in a way that minimizes some per-letter metric $d(r,r')$ for ratings.
To provide further insights into the proposed factor graph model and
the IMP algorithm, we compared our results against three other algorithms:
OptSpace \cite{Keshavan-arxiv09b}, SET \cite{Dai-arxiv09} and SVT
\cite{Cai-arxiv08}. Due to time and space constraints, we have chosen
three algorithms among all the available algorithms. OptSpace and
the more recent SET appear to be the best (this is also apparent from
experimental results), and can handle reasonably large matrix sizes.
In some cases, the programs are publicly available (e.g., \cite{Keshavan-arxiv09b}\cite{Cai-arxiv08})
and others (e.g., \cite{Dai-arxiv09}) have been obtained from their
respective authors. Our program is also publicly available at \cite{IMP-web}.

To make a fair comparison between different algorithms/models whose
complexity varies widely, we have created two smaller submatrices
from the real Netflix dataset:
\begin{itemize}
\item \textbf{Netflix Data Matrix 1} is a matrix given by the first 5,000
movies and users. This matrix contains 280,714 user/movie pairs. Over
15\% of the users and 43\% of the movies have less than 3 ratings. 
\item \textbf{Netflix Data Matrix 2} is a matrix of 5,035 movies and 5,017
users by selecting some 5,300 movies and 7,250 users and avoiding
movies and users with less than 3 ratings. This matrix contains 454,218
user/movie pairs. Over 16\% of the users and 41\% of the movies have
less than 10 ratings. 
\end{itemize}
Also, we hide 1,000 randomly selected user/movie entries as a validation
set $S$. The performance is evaluated using the root mean squared
error (RMSE) of prediction on this set defined by \[
\sqrt{\sum_{(n,m)\in S}\left(\hat{r}_{n,m}-r_{n,m}\right)^{2}/\left|S\right|}.\]
We primarily focused on the RMSE as a function of the average number
of observation ratings per user (i.e., how many ratings, $|O|$, are
needed to get each algorithm in shape). Simulations were performed
in the very small sample regime (e.g., much less than 0.5\% of ratings)
by varying the randomly selected average number of observed ratings
per user between 1 and 30 and the average results are shown in Figure
\ref{fig:result}. Note that the choice of parameters for each algorithm
(e.g., $g_{u}$ and $g_{v}$ for IMP and rank for others) was optimized
over the validation set $S$ by running each algorithm multiple times.
For IMP, we used hard K-means clustering (i.e., soft K-means clustering
with large $\beta$) for Algorithm \ref{alg:CR_alg} Step III to improve
the speed of $w(r|u,v)$ initialization. Also, to make a fair comparison
with algorithms that provide unbounded predictions, we clip the out-of-range
predictions (i.e., ratings greater than 5 or less than 1), if there
are any. \vspace{-1mm}

\subsection{Discussion}

\begin{figure*}[t]
\begin{centering}
\includegraphics[scale=0.47]{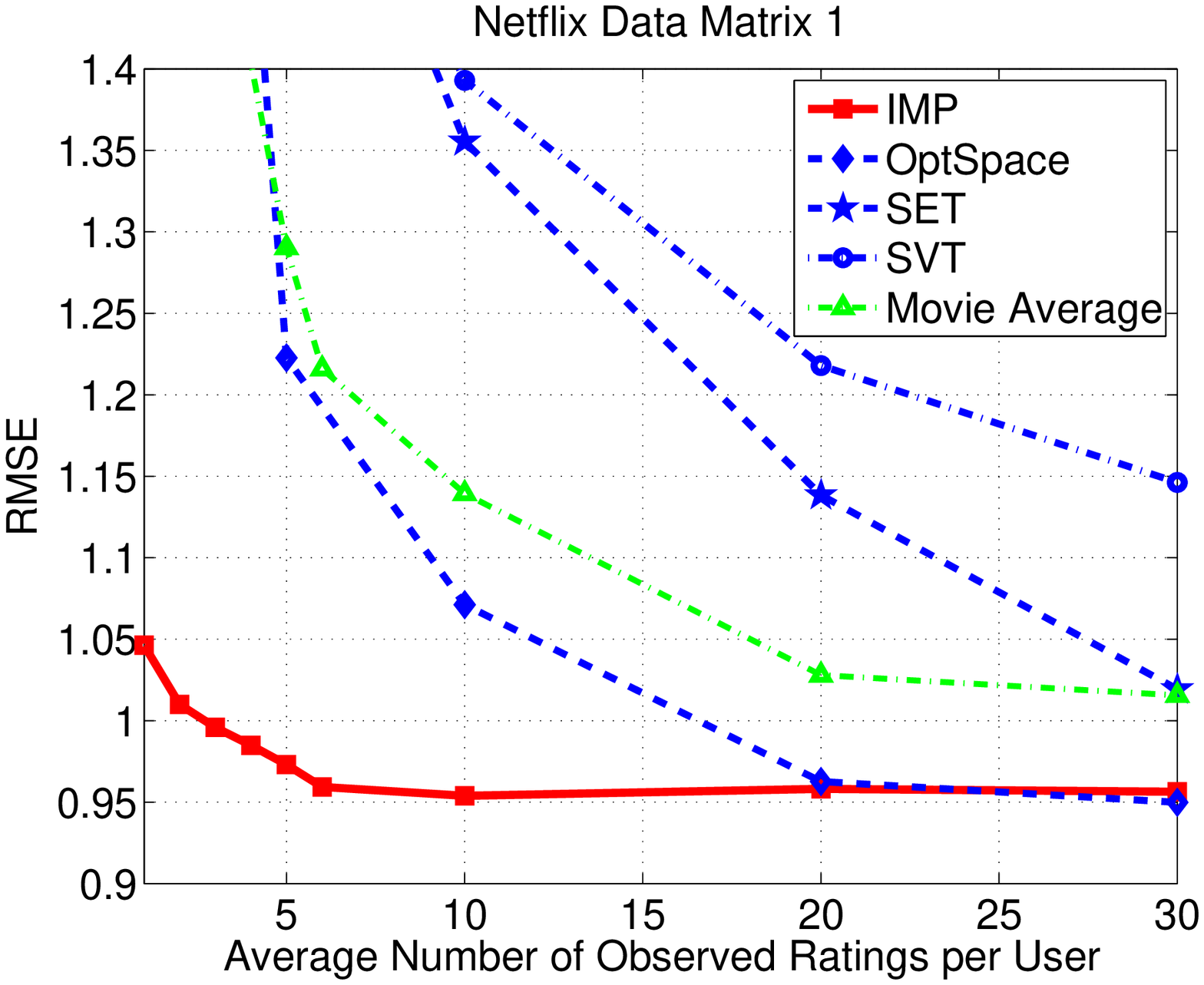}~~~\includegraphics[scale=0.47]{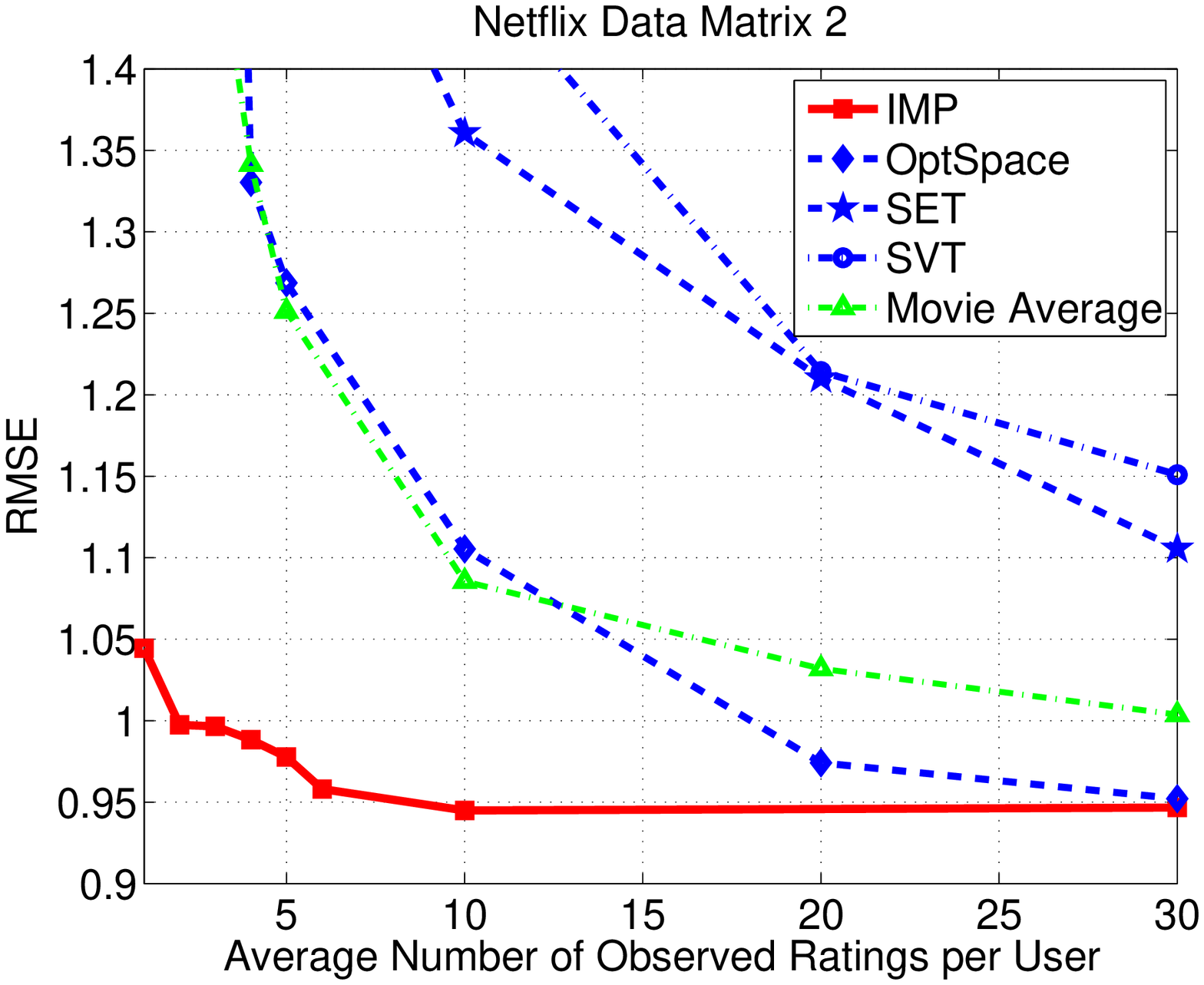}\vspace{-2mm}
\par\end{centering}

\caption{\label{fig:result}Remedy for the Cold-Start Problem: RMSE performance
is compared with other different competing algorithms \cite{Keshavan-arxiv09b}\cite{Dai-arxiv09}\cite{Cai-arxiv08}
on the validation set versus the average number of observations per
user for Netflix sub matrices.}
\vspace{-2mm}
\end{figure*}
Our results do shed some light on the performance of recommender systems
based on the MP framework. First, we have verified that IMP really
does improve the cold-start problem. From simulation results on Netflix
submatrices in Figure \ref{fig:result}, we clearly see while other
matrix completion algorithms perform similarly with large amounts
of revealed entries, the IMP algorithm can estimate the matrix very
well only after a few observed entries. The performance of other algorithms
for users with fewer than 5 ratings is generally poorer than that
of the simple movie average algorithm that uses the average rating
for each movie as the prediction. The IMP algorithm, however, performs
considerably better on users with a very few ratings. This better
threshold performance (see the steep RMSE decay) of the IMP algorithm
in comparison to other algorithms helps to reduce the cold start problem.
It is worth noting that the simple K-means clustering (used for $w(r|u,v)$
initialization) performs worse than movie average in the small sample
regime (due to space limits, this curve is not shown). This implies
that the improvement of IMP for the cold start problem comes from
the MP update steps and not the clustering initialization. We believe
this will be a major benefit of MP approaches to standard CF problems.
Other than these important advantages, each output group has generative
nature with explicit semantics. In other words, after learning the
density, we can use them to generate synthetic data with clear meanings.
These benefits do not extend to general low-rank matrix models easily.
\vspace{0mm}

\section{Conclusions \label{sec:Concl}\vspace{0mm}}

This paper introduces a novel MP framework for the matrix completion
problem associated with recommender systems. In contrast to prior
work, we model the problem using a generative factor graph model.
Based on the model, we introduce the IMP algorithm, which is a low
complexity inference method that gives optimal performance when the
graph is tree. We demonstrate the superiority of the IMP algorithm
by the comparing results against three other algorithms. Simulations
are performed with the focus on the cold-start setting (very sparse
regime) using Netflix data submatrices. Results show that, while the
methods perform similarly with large amounts of data, the IMP algorithm
is superior for very small amounts of data and improves the cold-start
problem for CF systems in practice. Another advantage of the IMP algorithm
is that it can be analyzed using the technique of DE that was originally
developed for MP decoding of error-correcting codes. We anticipate
that, by including the effects of clustering, this analysis will help
us understand the algorithm's impressive performance. \vspace{0mm}

\bibliographystyle{IEEEtran}

\end{document}